\documentclass[doublecol]{epl2}

\usepackage[greek,english]{babel}
 \usepackage{graphicx}
  
 \RequirePackage{color}

 \definecolor{MyDarkGreen}{rgb}{0.02,0.60,0.06}

\usepackage{teubner}

\def\mathcoppa{\hbox{\foreignlanguage{greek}{\coppa}}}
\def\smallcoppa{{\hbox{\foreignlanguage{greek}{\footnotesize\coppa}}}}
\def\oureta{{{\eta_Q}}}
\def\ourG{{G_Q}}
\def\ourD{{D_Q}}
\def\ouretahat{{\hat{{\eta}}_Q}}

\title{Fisher's scaling relation above the upper critical dimension}
\shorttitle{Fisher's scaling relation above the upper critical dimension} 

\author{R. Kenna\inst{1} \and B. Berche\inst{2}}

\institute{                    
  \inst{1} Applied Mathematics Research Centre, Coventry University, Coventry, CV1 5FB, England\\
  \inst{2} Statistical Physics Group, Institut Jean Lamour, UMR CNRS 7198,
Universit{\'{e}} de Lorraine, 
B.P. 70239, 54506 Vand\oe uvre l\`es Nancy Cedex, France
}


\abstract{Fisher's fluctuation-response relation is one of four famous scaling formulae and is consistent with  a vanishing correlation-function anomalous dimension above the upper critical dimension $d_c$.
However, it has long been known that numerical simulations deliver a negative value for the anomalous dimension there.
Here, the apparent discrepancy is attributed to a distinction between 
the system-length and correlation- or characteristic-length scales. 
On the latter scale, the anomalous dimension indeed vanishes above $d_c$ and Fisher's relation holds in its standard form.
However, on the scale of the system length, the anomalous dimension is negative and Fisher's relation requires modification.
Similar investigations at the upper critical dimension, where 
dangerous irrelevant variables become marginal, lead to an analogous pair of Fisher relations for logarithmic-correction exponents.
Implications of a similar distinction between length scales in percolation theory above $d_c$ and for the Ginzburg criterion are briefly discussed.
}
\begin{document}

\maketitle

\section{Introduction}

The scaling hypothesis was developed half a century ago and  stands as 
one of the pillars of modern theories of critical phenomena \cite{Stanley}.
In its basic form, six standard critical exponents are linked through four scaling relations \cite{history}.
Of these, hyperscaling and Fisher's fluctuation-response relation are notable in that the former involves the dimensionality $d$  and the latter involves the anomalous dimension $\eta$ \cite{Fi64}. 

Above $d=d_c$ dimensions, critical exponents assume their Landau values 
and hyperscaling was long considered to fail there  ({see} e.g., \cite{history,PrFi85,BDT,Ma,LeBellac}).
However, the introduction of a new, seventh, fundamental exponent ${\mathcoppa}$ (see footnote\footnote{
The notation $\alpha$, $\beta$, $\gamma$, $\delta$, $\eta$ and $\nu$ for the six primary critical exponents was standardised by Michael~E.~Fisher in the 1960's. 
In Ref.~\cite{BeKe12}, we introduced the new exponent  $q$ as characterising the leading, power-law FSS of the correlation length, in analogy to the exponent $\hat{q}$, introduced  in Ref.~\cite{KJJ2006} which characterises the logarithmic-correction term there.
Here we follow a suggestion by Fisher to switch  to the archaic Greek  letter $\mathcoppa$ (``koppa'' or ``qoppa'' -- the source of Latin ``q'') to synchronise more closely with his standard  nomenclature. We are grateful for this suggestion.
}) extends hyperscaling beyond the upper-critical dimension \cite{BeKe12}. 
From Fisher's dangerous-irrelevant-variables formalism, $\mathcoppa = d/d_c$ when $d>d_c$. 
Below the upper critical dimension, $\mathcoppa$ reverts to $1$.
Evidence that $\mathcoppa$ is {\emph{both physical and universal}}  was given in Ref.~\cite{BeKe12}. Physically, it is the exponent which governs the leading finite-size behaviour of the correlation length.
Its universality is evidenced by finite-size scaling (FSS) at the pseudocritical point
for both periodic and free boundary conditions \cite{BeKe12}.

The disentanglement of the correlation length from the actual length of the system when $\mathcoppa \ne 1$ is central to the extension of hyperscaling above $d_c$~\cite{BeKe12}. 
Length scales also enter via the correlation function into the definition of the anomalous dimension and derivation of Fisher's  relation \cite{Fi64}. 
Here we uncover  associated subtleties above $d_c$ which require re-interpretation of the formalism there. 
In particular, we show that the current paradigm, which does not distinguish length scales and their associate dimensionalities, violates bounds on the anomalous dimension, violates Fisher's scaling relation, and leads to disparities between field-theoretic results in the thermodynamic limit and numerical simulations in finite volume.
Our new formalism, which resolves all of these anomalies, entails {\emph{two}} anomalous dimensions and {\emph{two}} Fisher relations -- one for each length scale.
Additionally, the theory delivers {\it{two}} logarithmic analogues to the anomalous dimension and {\it{two}} corresponding relations {\it{at}} the upper critical dimension.

\section{Background: The Scaling Paradigm}
We follow the notation of Ref.~\cite{BeKe12} and denote by $P_L(t)$ the value of a function $P$ for a system of linear extent $L$ at reduced temperature $t = |T/T_L-1|$, where $T_L$ is the pseudocritical value of the temperature  $T$ 
(e.g., 
defined as the location of the susceptibility peak in vanishing external field) 
and  approaches the critical value $T_c$ as $L \rightarrow \infty$. 
The leading scaling behavior for the specific heat, spontaneous magnetization,  susceptibility and correlation length
are
\begin{equation}
 c_\infty(t) \sim  t^{-\alpha}, \,
 m_\infty(t) \sim  t^{\beta}, \,
 \chi_\infty(t) \sim  t^{-\gamma}, \,
 \xi_\infty(t) \sim  t^{-\nu},
\label{scaling}
\end{equation}
respectively. 
The correlation function is usually written as
$
 G(t,r) \sim {r^{-p}}{
                      D{\left[{
                                   {r}/{\xi_\infty(t)}
                                  }\right]
                           }
                }
$, 
for a function $D$ which, for $r \gg \xi_\infty(t)$, decays exponentially,
$
 D(y) \sim {
                       \exp{\left({
                                   -y
                                  }\right)
                           }
                }
$.
When $r \ll \xi_\infty(t)$ the correlation function reduces to
\begin{equation}
 G(t,r) \sim r^{-(d-2+\eta)},
 \label{G2}
\end{equation}
to leading order.
Above $d_c$, mean-field (MF) exponents describe scaling behaviour, and
for the Ising model  and associated $\phi^4$ theory, for which $d_c=4$, these exponents are
$\alpha =  0$, $\beta  = 1/2$, $\gamma  = 1$, $\delta = 3$,
$\eta = 0$, $\nu  = 1/2$.
The standard hyperscaling relation,
\begin{equation}
 \nu d = 2 - \alpha ,
\label{hyperscaling}
\end{equation}
fails for $d>d_c$.

Eq.(\ref{scaling}) may be expressed in terms of correlation length, e.g., 
$\chi_\infty(t) \sim \xi_\infty(t)^{\gamma/\nu}$.
Since standard FSS is controlled by the ratio of the correlation length to the actual length, 
the replacement $ \xi_\infty(t) \rightarrow \xi_L(0) \sim L$ then delivers the standard FSS formulae
\cite{PrFi85}
\begin{equation}
 c_L(0) \sim  L^{{\alpha}/{\nu}}, \quad
 m_L(0) \sim  L^{-{\beta}/{\nu}}, \quad
 \chi_L(0) \sim  L^{{\gamma}/{\nu}}.
\label{FSS1}
\end{equation}

Recently a seventh exponent was introduced which characterises the FSS of the correlation length above, as well as below, $d_c$~\cite{BeKe12}
(see also Ref.~\cite{ourCMP}),
\begin{eqnarray}
 \xi_L (0) \sim L^{\smallcoppa}
 \quad {\mbox{where}} \quad 
  {\mathcoppa} & = & \left\{{\begin{array}{l}
                                             ~~ 1,  \quad  {\mbox{if}} \quad d \le d_c  \\
                                             ~ {d}/{d_c}, \quad  {\mbox{if}} \quad d \ge d_c.
                                             
                             \end{array}}\right.
 \label{q}
 \end{eqnarray}
This seventh exponent originates in Fisher's dangerous-irrelevant-variable mechanism \cite{FiHa83}
provided an earlier assumption \cite{BNPY}  that the  finite-size correlation length $\xi_L$ is bounded by the length $L$ is relaxed.  
In Ref.~\cite{BDT}, $\xi_L$ was referred to as {\it{a characteristic length}}.
That it is, in fact, the finite-size {\it{correlation length}} was established directly in Refs.~\cite{BeKe12,JoYo05} for periodic boundary conditions and indirectly in    Ref.~\cite{BeKe12} for free boundaries.
The  exponent $\mathcoppa$ extends hyperscaling and FSS beyond the upper critical dimension 
via the relation
\begin{equation}
  {\nu d}/{{\mathcoppa}} = 2 - \alpha , 
\label{hyperhyperscaling}
\end{equation} 
and is supported analytically \cite{Br82,BeKe12} and numerically \cite{JoYo05,BeKe12}. 

Instead of being governed by the ratio of two length scales $\xi_\infty/L$, FSS now emerges 
through the ratio of the correlation volume in $d_c$ dimensions to actual volume, namely 
${\xi_\infty^{d_c}}/{L^d}$. In other words the usual  prescription  is replaced by
$ \xi_\infty \rightarrow \xi_L = L^{\smallcoppa}$, which, from Eq.(\ref{scaling}) delivers
\begin{equation}
 c_L(0) \sim  L^{{\smallcoppa\alpha}/{\nu}}, \,
 m_L(0) \sim  L^{-{\smallcoppa\beta}/{\nu}}, \,
 \chi_L(0) \sim  L^{{\smallcoppa\gamma}/{\nu}}.
\label{QFSS2}
\end{equation}
Eq.(\ref{QFSS2}), termed $Q$-FSS in Ref.~\cite{BeKe12} to compactly distinguish it from Eq.(\ref{FSS1}), has been verified for systems with periodic  and  free boundary conditions \cite{JoYo05,Turks,BeKe12}.

Here we turn our attention to Fisher's fluctuation-response relation.
The standard derivation starts from the fluctuation-dissipation theorem, viz.
\begin{equation}
 \chi_L (t) \sim \int_a^{L}{{G}(t,r) r^{d-1}d r} .
\label{integral}
\end{equation}
Here $a$ is the lattice constant in condensed matter,  vanishing in the continuum field theory.
Close to criticality, where $t$ is sufficiently small, so that $r \ll \xi$,
\begin{equation}
G(t,r) \sim r^{-(d-2+\eta)} D\left[{{r}/{\xi(t)}}\right],
\label{Fsaid}
\end{equation}
and in the thermodynamic limit, Eq.(\ref{integral}) becomes
\begin{equation}
 \chi_\infty(t) \sim  \int_a^{\infty}{G(t,r)r^{d-1}dr} .
\end{equation}
We partition this as
\begin{equation}
 \chi_\infty(t) \sim  
 \int_a^{S\xi_\infty(t)}{D\left[{\frac{r}{\xi_\infty(t)}}\right] \frac{dr}{r^{\eta-1}}} 
 +
 \int_{S\xi_\infty(t)}^{\infty}{G(t,r)r^{d-1}dr}, 
\end{equation}
where $S$ is a constant.
The second term is assumed to give rise to additive corrections close to criticality, where $\xi_{\infty}(t)$ diverges. The first term gives
\begin{equation}
 \chi_\infty(t) \sim \xi_\infty^{2-\eta} (t) \int_{a/\xi_\infty (t)}^{S}{{D}(y) y^{1-\eta}d y}. 
\end{equation}
One assumes that the lower integral limit only contributes to additive corrections to scaling, yielding to leading order,
\begin{equation}
 \chi_\infty(t) \sim \xi_\infty^{2-\eta}(t).
\label{integrall}
\end{equation}
Eq.(\ref{scaling}) then gives Fisher's relation \cite{Fi64},
\begin{equation}
 \eta = 2 - {\gamma}/{\nu}.
 \label{F1}
\end{equation}
If $L$ is finite, on the other hand, a similar procedure gives (setting $a=0$ to extract the leading scaling)
\begin{equation}
 \chi_L(t) \sim \xi_L^{2-\eta} (t) \int_{0}^{S}{{D}(y) y^{1-\eta}d y},
\label{integral3}
\end{equation}
where $S = L/\xi_L(t)$. 
Provided $\xi_L(0) \sim L$, the standard FSS formulae (\ref{FSS1}) deliver 
$\chi_L(0) \sim \xi_L^{2-\eta}(0)$, which again recovers Fisher's relation.

\section{Inconsistencies Above the Upper Critical Dimension}

The above derivation of Fisher's scaling relation for finite-size systems runs into trouble if $d>d_c$. There, with $\xi_L(0) \sim L^{\smallcoppa}$, the 
upper integral limit in Eq.(\ref{integral3}) has a leading $L$-dependency, destroying the derivation {\it{even with the $Q$-FSS form for $\chi_L$.}}

\begin{figure}[t]
\begin{center}
\includegraphics[width=0.9\columnwidth, angle=0]{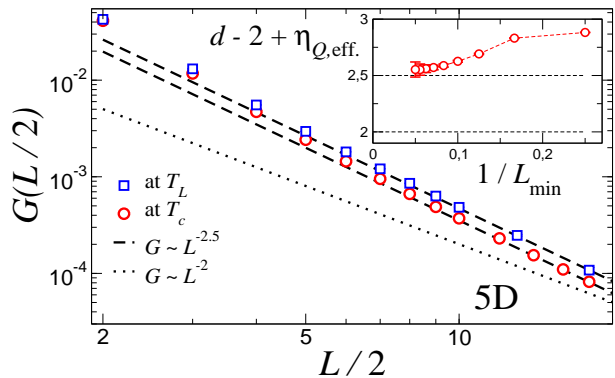}
\caption{The power-law decay of the correlation function for the $d=5$, critical and pseudocritical Ising model favours $\ourG (t,L/2) \sim L^{-5/2}$.
The insert shows that the effective exponent approaches the $Q$-theoretic value 
$\oureta = -1/2$ as the minimum lattice size used in the fit $L_{\rm{min}}$ increases.
}\label{5DG}
\end{center}
\end{figure}

To investigate further, we simulated the $d=5$
Ising model for periodic lattices \cite{BeKe12}.
Denoting the Ising spin at site $i$ of the lattice by $S_i$, 
the correlation function is 
$
G(t,i)=\langle{S_0S_i}\rangle-\langle{S_0}\rangle^2.
$
To extract the exponent $p$ from the general form 
${\displaystyle{G(t,r) \sim D(r/L)r^{-p}}}$ at criticality,  
$G(t,L/2)$ is plotted  against $L$ in  Fig.~\ref{5DG} \cite{LuBl}.
The result clearly supports $p= 5/2$.
If, as the standard paradigm purports, $p=d-2+\eta$, this would correspond to a value of the anomalous dimension of $-1/2$.

A \emph{negative} value for the anomalous dimension poses  problems.
Firstly it is in disagreement with mean-field theory and Landau theory, which deliver $\eta = 0$. Secondly, it violates Fisher's scaling relation (\ref{F1}). Thirdly, it  appears to violate  field theory which delivers a non-negative anomalous dimension for  a second-order phase transition if the underlying theory is of the 
Ginzburg-Landau-Wilson $\phi^4$ type~\cite{Fi64,Fi69,Delamotte}.

\section{How the Standard Paradigm Addresses the Problem of the Negative Anomalous Dimension}

The problem of the negative anomalous dimension was already noticed by Nagle and Bonner over 40 years ago \cite{NaBo70}.
In a numerical study, they determined the correlation decay in a spin chain with long-range interactions  and measured an anomalous dimension different from the standard one. 
In an attempt to explain this, Baker and Golner analytically determined spin-spin correlations in an Ising model for which scaling is exact \cite{BaGo73}.
Their explanation was that ``long long-range order'' is controlled by 
a {\emph{different}} anomalous dimension to the standard one, which controls ``short long-range order''. They found that the long long-range exponent  {\it{fails to satisfy the scaling relation for the anomalous dimension above the upper critical dimension.}}

The problem was revisited over a decade ago in a series of papers by Luijten and Bl{\"{o}}te \cite{LuBl}.
To recount their analysis,
 we again follow Fisher and first write the scaling form for the free energy density as \cite{FiHa83},
\begin{equation}
f_L(t,h,u) = b^{-d} f_{L/b} \left({tb^{y_t},hb^{y_h},ub^{y_u}}\right),
\label{BL4.2} 
\end{equation}
where $u$ is the scaling field associated with the coefficient of the quartic term in the Landau expansion.
Above $d_c$, the critical behaviour is controlled by the Gaussian fixed point in the renormalization-group formalism, where \cite{Ma}
\begin{equation}
 y_t = 2, \quad y_h = 1+{d}/{2}, \quad y_u = 4-d.
\end{equation}
Because of a discrepancy between $\alpha$, $\beta$ and $\delta$ coming from directly
differentiating Eq.(\ref{BL4.2}) and MF estimates, Fisher introduced {the notion of}
dangerous irrelevant variables for the free energy density 
in the thermodynamic limit  \cite{FiHa83}.
When $u \rightarrow 0$, Eq.(\ref{BL4.2}) becomes \cite{BNPY}
\begin{equation}
f_L(t,h,u) = b^{-d} f_{L/b} \left({tb^{y_t^*},hb^{y_h^*}}\right)
= L^{-d} f_{1} \left({tL^{y_t^*},hL^{y_h^*}}\right),
\label{BNPY}
\end{equation}
where
$
 y_t^* = y_t+p_2y_u$ 
 and $ y_h^* = y_h + p_3 y_u
$.
Landau exponents are recovered if $p_2= -1/2$ and $p_3=-1/4$ \cite{BNPY,BZJ85}.

No similar dangerous-irrelevant-variable mechanism  was expected for the correlation length or correlation function, since MF theory  and Gaussian fixed-point values of $\gamma$, $\eta$ and $\nu$, which are all connected to the correlation function, agree.
Notwithstanding this, similar considerations for the correlation length deliver 
\begin{equation}
\xi_L(t,h,u) = L^{\smallcoppa} \Xi \left({tL^{y_t^{*}},h L^{y_h^{*}}}\right).
\label{BNPY2}
\end{equation} 
In Ref.~\cite{BNPY}, $\mathcoppa$ was set to $1$ in Eq.(\ref{BNPY2})
 because of an expectation that $\xi_L$ is bounded by $L$.
For this reason, another length scale, 
 $\ell_\infty \sim t^{-1/y_t^{*}}$, was introduced in such a way that the first argument on the right-hand side of Eq.(\ref{BNPY2}) involves a ratio $\ell_\infty (t) / L$ which governs FSS. 
See also Ref.~\cite{Bi85,BDT}.

Luijten and Bl{\"{o}}te obtained the FSS of the correlation function by differentiating Eq.(\ref{BNPY}) with respect to two local magnetic fields $h(0)$ and $h(r)$ \cite{LuBl}.
When dangerous irrelevant variables are (incorrectly) not accounted for, $y_t^*$ and $y_h^*$ are replaced by $y_t$ and $y_h$, respectively, so that
$G \propto L^{2(y_h-d)} = L^{-(d-2)}$, which is the standard, Landau, MF result with $\eta = 0$.
However, (correctly) taking account of the dangerous irrelevancy in Eq.(\ref{BNPY}), Luijten and Bl{\"{o}}te obtained instead $G \propto L^{2(y_h^*-d)} = L^{-d/2}$, 
corresponding to an anomalous dimension  $\eta^* \equiv 2-d/2$.

Luijten and Bl{\"{o}}te give a second interpretation to their anomalous dimensions \cite{LuBl}. 
Writing the Ginzburg-Landau-Wilson action in momentum space, 
\begin{equation}
 F[\phi]
 = 
 \frac{1}{2}
 \sum_k{(k^2+\xi^{-2})|\phi_k|^2}
  + \frac{u}{4L^d}\sum_{k_1,k_2,k_3}{\phi_{k_1}\phi_{k_2}\phi_{k_3}\phi_{k_4}}
\label{BL19973}
\end{equation}
where $k_4=-k_1-k_2-k_3$.
Ignoring the danger by setting $u$ to zero, the  correlation function is identified as the inverse of the quadratic part of the action, leading again to the Ornstein-Zernike expression,
\begin{equation}
 G^{-1}(t,k) = k^2+\xi^{-2}(t).
\label{OZ}
\end{equation}
From the general form $G^{-1}(t,k) = k^{2-\eta}+\xi^{-2}(t)$, one identifies the Gaussian value $\eta = 0$.
The same result is obtained from Eq.(\ref{BL19973}) by first taking the thermodynamic limit $L \rightarrow \infty$.

Keeping the quartic term in Eq.(\ref{BL19973}) with finite $L$, however, the full quadratic part is 
\begin{equation}
 \frac{1}{2}
 \sum_k{\left({k^2+\xi^{-2} + \frac{3u}{2L^d}\phi_0^2}\right)|\phi_k|^2},
\label{OZ1}
\end{equation}
where $\phi_0$ is the zero mode associated with periodic boundary conditions.
Since $\langle{\phi_0^2}\rangle$ behaves as $\chi_L\sim L^{d/2}$, 
the final term in parentheses is $L^{-d}\phi_0^2 \sim L^{-d/2}$, which we identify as 
$k_{\rm{min}}^{d/2}$ acting as an additional momentum term.
It was argued in Refs.~\cite{LuBl} that this term dominates large distance behaviour 
in Eq.(\ref{OZ1}), leading to $\eta^* = 2-d/2$.

To summarise, in the standard paradigm there is a discrepancy between the 
Landau MF value $\eta = 0$ for the anomalous dimension above $d_c$ and the 
value $\eta^* = 2 - d/2$ measured on finite systems. 
On the one hand this discrepancy is linked to neglecting, or accounting for, the dangerous
irrelevant variable $u$ (leading to $\eta = 0$ or $\eta^* = 2-d/2$, respectively).
On the other hand it is attributed to a difference between short long-range ($\eta = 0$) and long long-range behaviour ($\eta^* = 2-d/2$). 

The standard paradigm does not, however, explain how $\eta^* = 2-d/2$ for long long distance is manifest as $\eta=0$ in the infinite-volume limit where  field-theoretic theorems outlawing negative anomalous dimensions apply.
Nor does it explain why it is the   correct, dangerous-irrelevant-variables, long long-range $\eta^*$ which conflicts with Landau and MF theory, 
fails to satisfy Fisher's relation and violates field theory.
(One would rather expect the conflict to be associated with the  incorrect processes of neglecting dangerous irrelevant variables or taking short rather than long  long distances.)
Therefore {\emph{the standard paradigm does not explain scaling above the upper critical dimension.}} 

\section{Resolution of Puzzle}

Here we offer an alternative explanation for the negativity of the measured value of the anomalous dimension, based on the $Q$-theory proposed in Ref.~\cite{BeKe12,ourCMP}.
This new explanation also resolves all of the above puzzles.
According to the theory, there is a difference between the underlying length scale $L$ of the system above $d_c$ and its correlation length scale $\xi_L$. 
This difference is manifest as $\xi_L \sim L^{\smallcoppa}$.

In Eq.(\ref{Fsaid}), the distance $r$ is implicitly measured on the correlation length scale and this leads to the usual Fisher relation (\ref{F1}).
In Eq.(\ref{integral3}), however, the length scales $L$ and $\xi_L$ are incorrectly mixed
above the upper critical dimension.

To repair this, we write the critical correlation function in terms of the system-length scale  as
\begin{equation}
\ourG(0,r) \sim r^{-(d-2+\oureta)} \ourD \left({{r}/{L}}\right),
\label{GscaleL}
\end{equation}
where $\oureta$ is the anomalous dimension measured on this scale,
the subscript indicating that $Q$-FSS (\ref{QFSS2}) 
rather than standard FSS (\ref{FSS1})
prevails there \cite{BeKe12}.

Eq.(\ref{integral3}) for the susceptibility is then
\begin{equation}
 \chi_L(0) \sim \int_0^L{\!\!r^{1-\oureta} \ourD \left({\frac{r}{L}}\right) dr} = 
 L^{2-\oureta} \!\!\! \int_{0}^{1}{{\!\!\ourD}(y) y^{1-\oureta}d y}.
\label{integral4}
\end{equation}
Above $d=d_c$, the $Q$-FSS formulae (\ref{QFSS2}) then yield
\begin{equation}
  \oureta= 2 - \mathcoppa \gamma / \nu.
\label{QFisher}
\end{equation}
In the Ising case, where $\mathcoppa = d/4$, $\gamma = 1$ and $\nu = 1/2$, 
this gives $\oureta = 2-d/2$ and identifies $\oureta$ with  $\eta^*$ of Refs.\cite{LuBl,NaBo70,BaGo73}. 
Eq.(\ref{QFisher}) is the fluctuation-response relation above the upper critical dimension when distance is measured
on the scale of system size.
The standard expression (\ref{F1}) is the equivalent formula there   when distance is measured
on the correlation-length scale.
The relationship between the two anomalous dimensions is then
\begin{equation}
 \oureta = \mathcoppa\eta + 2(1-\mathcoppa).
\label{etaetaQ}
\end{equation}
Below $d_c$, the two anomalous dimensions coincide. 
When $d>d_c$, $\oureta$ is negative.
Since the non-negativity bounds for  the anomalous dimension 
refer to correlation decay on the scale $\xi$, they involve $\eta$ rather than $\oureta$, and are not violated~\cite{Fi64,Fi69,Ma,LeBellac,Delamotte,ZJ}.

This interpretation advocates that there are two forms for the correlation function, two anomalous dimensions and two Fisher  relations, depending on whether distance is measured on the scale of $L$ or $\xi_L \sim L^{\smallcoppa}$.
The value $\eta = 0$ is correct when distance is measured on the scale of $\xi_L$ and $\oureta = 2-d/2$ is correct on the length-scale $L$. 
Both are  valid as characterising long-distance decay.
On either scale, there is no need to distinguish between short long distances and long long distances.
Our numerical results for the short-range model, and Luijten's and Bl{\"{o}}te's numerics for its long-range counterpart, confirm $\oureta$ or $\eta^* = 2-d/2$ as governing the correlation decay at criticality \cite{LuBl}.

\section{Logarithmic Corrections at $d_c$}

Thus, there is no numerical disagreement between our results and those of Refs.~\cite{LuBl,NaBo70,BaGo73} for $d>d_c$.
At this point neither theory is falsefied by numerics.
Instead, interpretations differ. But these interpretations are important at a fundamental level. While each interpretation can be couched in terms of the dangerous-irrelevant-variables mechanism, the paradigm hitherto relies completely on the role of the quartic term.
To discriminate between them, we need a scenario without dangerous irrelevant variables and $d=d_c$  presents  such a case.
There $u$ is marginal with logarithmic corrections arising from the renormalization-group formalism. 
We shall now show that, while the Luijten-Bl{\"{o}}te scenario has no consequence at $d=d_c$,
our scaling theory again leads to two correlation functions and to logarithmic analogues to each of the Fisher relations (\ref{F1}) and (\ref{QFisher}). 
This  provides a route to test interpretations  numerically.

We follow the notation of Ref.~\cite{KJJ2006} and denote the logarithmic-correction 
exponents, that are known to appear at $d=d_c$,  by hatted indices, 
\begin{eqnarray}
   \chi_\infty(t)  & \sim & t^{-\gamma} |\ln{t}|^{\hat{\gamma}} ,
   \label{a1}\\
   {\xi}_\infty(t) & \sim & t^{-\nu}    |\ln{t}|^{\hat{\nu}} ,
   \label{a2}\\
   {\xi}_L(0)      & \sim & L (\ln{L})^{\hat{\smallcoppa}} .
   \label{qhat}
\end{eqnarray}
For the Ising and $\phi^4$ models, $\hat{\gamma} = 1/3$, $\hat{\nu} = 1/6$, $\hat{\mathcoppa}=1/4$ \cite{Br82,KJJ2006}.
(The exponent $\hat{\mathcoppa}$ was written $\hat{q}$ in Ref.~\cite{KJJ2006}.)
The correlation function in the critical region is
\begin{equation}
 G(0,r) \sim D  \left[{y(r)}\right]  r^{-(d-2+\eta)}(\ln{r})^{\hat{\eta}} ,
\label{Glog}
\end{equation}
where $y(r) = r/\xi_L(0)$ if distance is measured on the correlation-length scale.
If the system-length scale is used instead, then $y(r) = r/L$ and $\ouretahat$ 
replaces ${\hat{\eta}}$ in Eq.(\ref{Glog}).
For the $d=4$ Ising model  $\hat{\eta}=0$ \cite{KJJ2006}.
With logarithmic corrections, Eq.(\ref{integrall}) and Eq.(\ref{integral4}) become
\begin{eqnarray}
 \chi_\infty(t) & \sim & \xi_\infty^{2-\eta}(t)[\ln{\xi_\infty(t)}]^{\hat{\eta}}
 \left[{1+ {\mathcal{O}}(1/\ln{\xi_\infty})}\right],
 \\
 \chi_L(0) & \sim & L^{2-\oureta}(\ln{L})^\ouretahat
 \left[{1+ {\mathcal{O}}} (1/\ln{L})\right].
\end{eqnarray}
Inserting Eqs.(\ref{a1}) and (\ref{a2}) and their FSS counterparts, respectively, yields the analogues to Eqs.(\ref{F1}) and (\ref{QFisher}),
\begin{eqnarray}
 \hat{\gamma} & = & (2 - \eta) \hat{\nu} + \hat{\eta},
 \label{50}
 \\
 \hat{\gamma} & = & (2 - \oureta) (\hat{\nu}-\hat{\mathcoppa}) + \ouretahat.
 \label{61}
\end{eqnarray} 
Of course, $\oureta = \eta$ in Eqs.(\ref{Glog})-(\ref{61})
since $d=d_c$ there.
The relation (\ref{50})  is the same as that proposed in Ref.~\cite{KJJ2006}.
Indeed, in Ref.~\cite{KJJ2006}, this formula was verified in 
a variety of models at their respective 
upper critical dimensions in the infinite-volume limit, through exponential decay of the correlation function, i.e.,  where distance  is measured in units of the correlation length.

\begin{figure}[t]
\begin{center}
\includegraphics[width=0.9\columnwidth, angle=0]{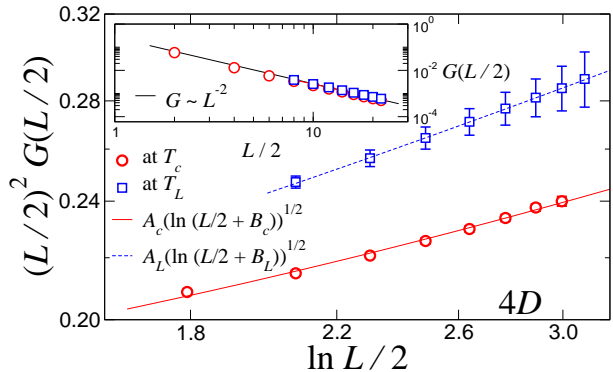}
\caption{In $d=4$ dimensions, the logarithmic corrections to $G(t,L/2)$ have a positive exponent, which
is compatible with the $Q$-theoretic $\ouretahat=1/2$ and incompatible with zero.
The insert shows that the leading singularity is governed by the exponent $d_c-2+{\eta} = 2$.}
\label{4DG}
\end{center}
\end{figure}

However, finite-size numerical approaches are defined on the underlying lattice
with length-scale $L$, for which 
\begin{equation}
 \ouretahat = \hat{\eta} + (2-\eta)  \hat{\mathcoppa} .
\label{etaqhat1o2}
\end{equation}
Thus $\ouretahat =1/2$ in the $d=4$ Ising model.
We test these predictions in Fig.~\ref{4DG} where $(L/2)^2 G(t,L/2)$ is plotted against  $\ln{(L/2)}$ at both the critical and pseudocritical points. 
The positive slope is clearly not $\hat{\eta}=0$. 
Compatibility with $\ouretahat=1/2$  is evident and fits to $A (\ln{(L/2 + B)})^{1/2}$, both at criticality and at pseudocriticality, 
are nicely compatible with the numerical data.

\section{Discussion}

Returning to $d>d_c$ case, our claim is that both $\eta$ and $\oureta$ are valid at long distances.
For this claim not to violate Fisher's dangerous-irrelevant-variables theory, $\eta$ should also arise from it, just as  $\oureta$ does.
Indeed we can see the emergence of both anomalous dimensions  through the scaling of the correlation function in a manner similar to the development of 
Eqs.(\ref{BNPY}) and (\ref{BNPY2}) above.
From dimensional analysis, one may write the standard form
\begin{equation}
 G_L(t, u,r) = b^{-2X_\phi} G_{L/b} \left({tb^{y_t},ub^{y_u},rb^{-1}}\right),
\label{BNPYus}
\end{equation}
in which $X_\phi = d/2-1$.
Note that $r$ and $b$ have the same dimension as $L$.
Acknowledging the danger of $u$, we treat this in a similar manner to 
Eqs.(\ref{BNPY}) and (\ref{BNPY2})
and write
\begin{equation}
 G_L(t,u,r) 
= b^{2-d+y_uv_1} \bar{G}_{L/b} \left({tb^{y_t^*}, rb^{-1+y_uv_2}}\right).
\label{BNPY3}
\end{equation}
Interpreting $r$ as a length requires $v_2=0$ to render the final argument on the right  dimensionless.
With $t=0$ and $b=r$, we then obtain $G_L(0,u,r) = 1/r^{d-2-y_uv_1}$, which 
accords with $\ourG$ in Eq.(\ref{GscaleL}) provided that $v_1 = -\oureta/y_u = -1/2$.
If, on the other hand, we interpret $r$ as a correlation length, the final argument is dimensionless if it is $rb^{-\smallcoppa}$. We then require $v_2 = (1-q)/y_u = 1/4$.
Again setting $t=0$, but now setting $b^{\smallcoppa}=r$, we obtain the scaling of the  correlation function as
 $G_L(0,u,r) = 1/r^{(d-2-y_uv_1)/\smallcoppa}$.
Inserting $v_1=-1/2$ delivers the Ornstein-Zernike form $G(0,u,r) \sim 1/r^2$.


In conclusion, 
for a comprehensive picture of scaling above the upper critical dimension, one must take care whether distance is measured in terms of the system-length scale or the correlation-length 
scale.
To track these, two correlation functions are required, resulting in two Fisher relations, involving two anomalous dimensions, 
{\emph{only one of which  is captured by Landau theory and MF theory.}}
The hidden anomalous dimension is
revealed through numerical simulations on the system-length scale.
At the upper critical dimension itself, analogous expressions arise for the logarithmic corrections to scaling there.


The magnetisation transitions in spin models are equivalent to percolation transitions of Fortuin-Kasteleyn clusters.
The 30-year-old prevailing picture of hyperscaling breakdown in percolation theory predicts that the number of spanning clusters $N_L$ is finite for $d<d_c$ but diverges as $N_L \sim L^{d-d_c}$ for $d>d_c$ 
($d_c=6$ for  percolation theory) \cite{Conig85Co00}.
The  theory also predicts that the critical clusters have fractal dimension $D= (\beta + \gamma)/\nu$, which is independent of $d$ when $d > d_c$.
This perceived clear demarcation between $d<d_c$ and $d>d_c$ has been steadily undermined over the  years \cite{Hu96Hsu01,FoStCo03FoAhCoSt04}.
In Ref.~\cite{FoStCo03FoAhCoSt04}, because finite-size simulations did  not follow the standard theory, and $N_L \sim L^0$ is claimed instead 
depending on boundary conditions, the behaviour of $N_L$ above $d_c$ was declared an ``open issue''.

A simple thought experiment shows that the standard interpretation of $N_L \sim L^{d-d_c}$ spanning clusters is flawed. With interactions of sufficiently long range, one can construct a percolation or spin 
model with $d_c< 1$. That result then predicts a diverging number of spanning clusters in $d=1$ dimension despite there being only enough physical space to accomodate one such cluster there. There can, however, be a (finite) number of critical clusters of length $O(L)$.

The fundamental error undermining  prevailing percolation theory above $d_c$ is the assumption that $\xi_L \sim L$ ($\xi$ is the connectedness length in pure percolation theory). Using Eq.(\ref{q}) instead, carefully distinguishing finite $L$ from its infinite limit, and otherwise following Ref.~\cite{Conig85Co00}, one derives Eq.(\ref{hyperhyperscaling}) for all $d$. This approach delivers $N_L \sim L^0$, 
compatible with the above thought experiment and with the aforementioned claim in Ref.~\cite{FoStCo03FoAhCoSt04}. $Q$-theory also predicts that the mass of the critical clusters is $\xi_L^D = L^{D_Q}$ where $D_Q= \mathcoppa D$.
The fractal dimension of the critical clusters is therefore independent of $d$ only when measured on the correlation-length scale above $d_c$.

It is also legitimate to ask, in the present framework, about the status of the Ginzburg criterion, which defines $d_c$ as that dimension above which fluctuations become negligible and Landau exponents prevail.
It is usually obtained, for example, by comparing the fluctuations, measured by $\chi$,  with the average magnetization-squared both at the correlation-length scale. 
The standard argument is that, for MF theory to be correct, one should have $\chi \ll m^2 \xi^d$ or $d > d_c =  (\gamma + 2\beta)/\nu$. 
This defines $d_c=4$ when Landau exponents are used. 
Since we now know that above $d_c$ the correlation length exceeds the system size,  
the above argument is valid only at the scale $L$, where fluctuations $\chi_L\sim L^{\smallcoppa\gamma/\nu}$ now appear to be of the same order as the average square $m^2L^d\sim L^{d-2\smallcoppa\beta/\nu}$ when Landau
exponents are plugged in. 
This shows that the correlations have not been washed out at the size $L$ 
(but the correlators still decay as $r^{-d/2}$). Strictly speaking, MF theory is not fully valid above $d_c$: while the thermal 
exponents $\alpha$, $\beta$, $\gamma$ and $\nu$, and the magnetic counterpart $\delta$ are those of Landau theory, the exponent describing the space dependence of the correlation function is $\oureta$ rather than $\eta$, which describes an
emergent $4-$dimensional field theory at the scale of the correlation length.

\acknowledgments

We  thank M.E. Fisher, Yu.~Holovatch, F.~Igl{\'{o}}i and N.~Izmailian  for careful readings of the manuscript and helpful discussions.
We also thank M.E. Fisher for suggesting to introduce the symbol $\mathcoppa$ for the new exponent and J.~Cardy for advice on its typesetting.
We also thank F.~Igl{\'{o}}i for suggestion to include material on percolation theory and J.-C. Walter for help with the numerics 
This research was supported by 
Marie Curie  IIF and IRSES  
grants within 
the 7th 
EU Framework Programme.



\begin{thebibliography}{0}


\bibitem{Stanley}
H.E. Stanley,
Rev. Mod. Phys. {\bf{71}} (1999) S358-S366.
\bibitem{history}
M.E.~Fisher, 
Rev. Mod. Phys. {\bf{70}} (1998) 653-681.
\bibitem{Fi64}
M.E.~Fisher, 
J.~Math.~Phys. {\bf{5}} 944 (1964).
\bibitem{BDT}
I.G. Brankov, D.M. Danchev and N.S. Tonchev, 
Theory of critical phenomena in finite-size systems: scaling and 
quantum effects, 
 (World Scientific, Singapore, 2000).
\bibitem{Ma}
 S.-k.~Ma, 
 Modern Theory of Critical Phenomena,
 (Addison-Wesley, Redwood, CA, 1976).
\bibitem{LeBellac}
M.~Le~Bellac,
 Quantum and Statistical Field Theory,
 (Oxford Science Publications, Oxford, 1991).
\bibitem{PrFi85}
V.~Privman and M.E.~Fisher, J. Stat. Phys. 33 (1983) 385-417. 
\bibitem{BeKe12}
B.~Berche, R.~Kenna and J.-C.~Walter,
Nucl. Phys. B {\bf{865}} (2012) 115-132.
\bibitem{ourCMP}
R.~Kenna and B.~Berche,
Cond. Matter Phys. {\bf{16}} (2013) 23601.
\bibitem{FiHa83}
 M.E.~Fisher, 
 in Lecture notes in physics {\bf{186}}, critical phenomena, 
 ed F.J.W.~Hahne, 
 (Springer, Berlin, 1983)  pp. 1-139.
\bibitem{BNPY}
 K.~Binder,~M. Nauenberg, V.~Privman, and A.P.~Young, 
 Phys. Rev. B {\bf{31}} (1985) 1498-1502. 
\bibitem{JoYo05}
 J.L.~Jones and A.P.~Young,
 Phys. Rev. B {\bf{71}}  (2005) 174438.
\bibitem{Br82}
 E.~Br\'ezin, 
 J. Physique {\bf{43}} (1982)
 15-22.
\bibitem{Turks}
 N.~Aktekin, \c{S}~Erko\c{c} and M.~Kalay,
 Int. J. Mod. Phys. C {\bf{10}} (1999) 1237-1245;
 N.~Aktekin and {\c{S}}.~{Erko\c{c}},
 Physica A  {\bf{284}}  (2000) 206-214;
 Z.~Merdan, A.~Duran, D.~Atille, G.~M{\"{u}}lazimo{\u{g}}lu and A.~G{\"{u}}nen,
 Physica A   {\bf{366}}  (2006) 265-272.
\bibitem{LuBl}
 E.~Luijten and H.W.J.~Bl{\"{o}}te, 
 Phys. Rev. Lett. {\bf{76}} (1996) 1557-1561;
 Phys. Rev.~B {\bf{56}} (1997) 8945-8958;
 H.W.J.~Bl{\"{o}}te and E.~Luijten,
 Europhys. Lett. {\bf{38}} (1997) 565-570;
 E.~Luijten, 
 Interaction range, universality and the upper critical dimension, 
 (Delft University Press, Delft 1997).
\bibitem{Fi69}
 M.E.~Fisher,
 Phys. Rev. {\bf{180}} (1969) 594-600.
\bibitem{Delamotte}
B.~Delamotte,  D.~Mouhanna and M. Tissier, Phys. Rev. B {\bf{69}} (2004) 134413 (2004).
\bibitem{NaBo70}
J.F. Nagle and J.C. Bonner, 
J. Phys. C: Solid State Phys. {\bf{3}} (1970) 352.
\bibitem{BaGo73}
G.A. Baker, Jr. and G.R. Golner, 
Phys. Rev. Lett. {\bf{31}} (1973) 22-25.

\bibitem{BZJ85}
 E.~Br{\'{e}}zin and J.~Zinn-Justin,
 Nucl. Phys. B {\bf{257}}  (1985) 867-893.

\bibitem{Bi85}
 K.~Binder, 
 Z. Phys. B {\bf{61}} (1985) 13-23.


\bibitem{ZJ}
J.~Zinn-Justin, 
Quantum Field Theory and Critical Phenomena, 
(clarendon Press, Oxford, 2002).



\bibitem{KJJ2006} 
 R.~Kenna, D.A.~Johnston, and W.~Janke,
 Phys. Rev. Lett. {\bf 96}  (2006) 115701;
 ibid. {\bf 97} (2006) 155702.
 
 




\bibitem{Conig85Co00}
A. Coniglio, 
in Springer Proceedings in Physics, Vol. 5: Physics of Finely Divided Matter, 
by M.~Daoud, N.~Boccara (Eds.),
Springer, Berlin (1985) pp 84-101;
Physica A 281 (2000) 129-146.
\bibitem{Hu96Hsu01}
C.-K.~Hu and C.-Y.~Lin, Phys. Rev. Lett. {\bf{77}} (1996) 8-11;
H.-P.~Hsu, C.-Y.~Lin and C.-K.~Hu, Phys. Rev. E {\bf{64}} (2001) 016127.
\bibitem{FoStCo03FoAhCoSt04}
S. Fortunato, D.~Stauffer and A.~Coniglio,
Physica A {\bf{334}} (2004) 307-311;
S.~Fortunato, A.~Aharony, A.~Coniglio and D.~Stauffer,
Phys. Rev. E {\bf{70}} (2004) 056116.




\end{thebibliography}
\end{document}